# Evidence for Social Role in a Dolphin Social Network

David Lusseau[1,2]

[1]University of Otago, Department of Zoology, PO Box 56, Dunedin, New Zealand
[2]Present address: Dalhousie University, Department of Biology, 1355 Oxford Street, Halifax, NS B3H 4J1, Canada
E-mail: d.lusseau@dal.ca, Tel: (902) 494 3723

## Abstract

Social animals have to take into consideration the behaviour of conspecifics when making decisions to go by their daily lives. These decisions affect their fitness and there is therefore an evolutionary pressure to try making the right choices. In many instances individuals will make their own choices and the behaviour of the group will be a democratic integration of everyone's decision. However, in some instances it can be advantageous to follow the choice of a few individuals in the group if they have more information regarding the situation that has arisen. Here I provide early evidence that decisions about shifts in activity states in a population of bottlenose dolphin follow such a decision making process. This unshared consensus is mediated by a non-vocal signal which can be communicated globally within the dolphin school. These signals are emitted by individuals that tend to have more information about the behaviour of potential competitors because of their position in the social network. I hypothesise that this decision making process emerged from the social structure of the population and the need to maintain mixed-sex schools.

## Introduction

Individuals have to regularly make decisions that will affect their fitness (*1, 2*). In gregarious species these decisions are compounded by the need to interact with conspecifics in relation to intrinsic and extrinsic influences such as prey availability (*3-6*) or intra-specific competition (*7, 8*). These influences often result in the emergence of complex social behaviour (*9-11*) and social structure (*12-15*) which facilitate the decision-making process and often the need to reach consensus between individuals within a group (*16, 17*). Many social species have developed signals that help members of groups or aggregations to coordinate their activities (*18-20*). This decision process can results in conflicts (*2, 21*) and no consensus has been reached on which mechanisms are driving this process (*2, 17, 22, 23*). Conradt and collaborators discriminate between democratic and despotic decision-making processes and argue that democratic processes are widespread because they lower the consensus cost by producing less extreme decisions (*2*). However, it is hypothesised that despotic decision-making processes, i.e. a small subset of individuals influencing disproportionably the behaviour of the whole group, can emerge from situations where some individuals are more knowledgeable about the situation than others. In that case the cost of decision is lowered by following information holders (*2*).

The bottlenose dolphin (*Tursiops* sp.) population residing in Doubtful Sound, New Zealand (*24*) presents an interesting model to test this hypothesis. Bottlenose dolphins live in fission-fusion societies and therefore individuals can make choices to join or leave a group (*25*). The social relationships of all individuals in the population have been mapped (*26, 27*) using a network representation approach which helped in defining the affiliation 'distance' both between individuals and between clusters of individuals within the population (*27*). Two social units (communities hereafter) were identified in this population using these techniques (*27*).

Two behavioural events have been recently identified as signals of shifts in school's activity state (*28-30*). Side flopping (SF) is defined as a dolphin clearing its body entirely from the water and landing on its side and is only performed prior to the onset of travels. While upside-down lobtailing (ULT) is defined as a dolphin rolling to expose its ventral side at the water surface and slapping repeatedly the water surface with the dorsal side of its tail and is only observed before the school concludes travelling (*28-30*). These behaviours are therefore reminiscent of other signals used by a variety of species to synchronise activities (*18, 19*). While these signals are often used as a voting system in other species (*17*), it appears that SF and ULT are always performed by only one member of the school (*28, 30*). These events offer signals that can be emitted to all individuals within the school but which are not heard by non-school members because the sound produced by these percussive behaviours does not travel as far as vocalisations do (*31*).These two non-vocal behavioural signals could be advantageous to minimise the cost of intra-specific competition from direct competition for food between the different communities present in the population. They can help synchronising the activities of the school without advertising it to non-school members. There is no sign of complete segregation of social units in bottlenose dolphins (*13, 27, 32*). The social structure of the Doubtful Sound bottlenose dolphin population is such that a large proportion of individuals within schools (41%) spent a significant amount of time together, hence it would be advantageous to not only have a signal that would allow activity synchronisation but also would not allow this synchronisation to be advertised outside the school (*24*). If this was the case, individuals that spent some time with members outside their communities would be more likely to have knowledge about these potential competitors and hence be more reliable in decision making processes which involve competition evasion. Following their decisions would therefore lower the cost of choices and ultimately increase the fitness of school members (*2*).

Centrality measures on network graph can identify the location of individuals in relation with others (*27, 33, 34*). They can therefore help identifying individuals that have social relationships spread *between* clusters of individuals as well as individuals that have a more central position *within* these clusters. I therefore tested the likelihood that these non-vocal behavioural events were performed by individuals that were more likely to have a better knowledge of the activities of other clusters of individuals.

## Materials and methods

### *Field techniques*

I collected behavioural data in Doubtful Sound, New Zealand (45°30' S, 167°00' E) between June 2000 and May 2002. Systematic surveys of the fjord were conducted to look for dolphin schools (*24*). Once a school was detected the identity of individuals in the school was determined using photo-identification (*35*). A code of conduct was established for the observing vessel to minimise its effects on the focal schools (*36*). Studies showed that the behaviour of the focal schools was not affected by the presence of the observing vessel (*29, 37*). Side flops (SF) and upside-down lobtails (ULT) are rare events (0.012 sf/min and 0.016 ult/min of focal follows (*30*)); I therefore recorded the occurrence of side flops and upside-down lobtails in an *ad libitum* fashion while following the school (*38, 39*). Side flops were defined as jumps during which a dolphin cleared its entire body out of the water and landed on its side. Upside-down lobtails were defined as situations when a dolphin was upside-down stationary at the surface, belly pointing upwards, and forcefully slapped the water surface with its tail. Observations ended when the weather deteriorated, the focal school was lost, or the day ended, therefore the end of an observation period was not dependent on the behaviour of the focal school.

The gender of photo-identified individuals was assessed by direct observation of the genital slit using an underwater camera (*36*). Both the absence/presence of mammary slits and the distance between the genital and anal slits permitted to sex the animals (*36*). The identity of individuals performing the behavioural events was defined either through direct visual observations or from either photographs or videos. The marking rate in this population is high (*24, 40*) which means that practically all individuals can be recognised from marks on their dorsal fins. Therefore practically all the population (excluding calves) was equally likely to be recognised in this way, minimising sampling bias. Social relationships within the population have been previously described (*26, 27*) and this study is based on the same data which is based on school membership obtained using photo-identification (Figure 1). The resulting social network is defined by preferred companionships between individuals in the population (*26*).

### *Analytical techniques*

Centrality measures (degree and betweenness) for each individual present in the network were calculated using Ucinet (*41*). The higher the betweenness (*42*), the more often an individual is found between clusters in the network graph. In other words, betweenness quantifies how much of a bottleneck an individual is in the network. It is defined using shortest path length. For each possible pair of individuals in a network it is possible to find the shortest path to go from one to another by travelling along the edges of the network, passing from node to node. The betweenness of an individual (node) is measured by counting how often that node is frequented when travelling between all possible pairs using shortest paths. Individuals with high betweenness tend to be information brokers in human societies (*43*) and potentially in bottlenose dolphin societies as well (*27*). The degree of an individual (*44*) is a measure of how much influence an individual can have on its peers: the more individuals that a dolphin is linked to, the more individuals it has the opportunity to affect. The degree of an individual is measured by counting the number of associates a dolphin has (number of edges). There are early indications that these measures are behaviourally meaningful in dolphin societies as the temporary disappearance of individuals with high betweenness may have led to groups of individuals temporarily spending less time together (*27*). In addition, these centrality measures have proven useful to identify central individuals in other animal networks (*45-48*). Randomisation tests were used to compare the difference in average centrality measures between individuals that were observed performing the behaviours and others to average differences in which individuals were randomly assigned as behaviour performer or not.

Since SF and ULT are rare events, it is possible that an individual may not have been observed performing them because we did not spend enough time with it. To eliminate this potential sampling bias, the random selections within the randomisation tests were weighted by the amount of time we spent observing each individual in relation to the total amount of time we spent observing dolphin schools.

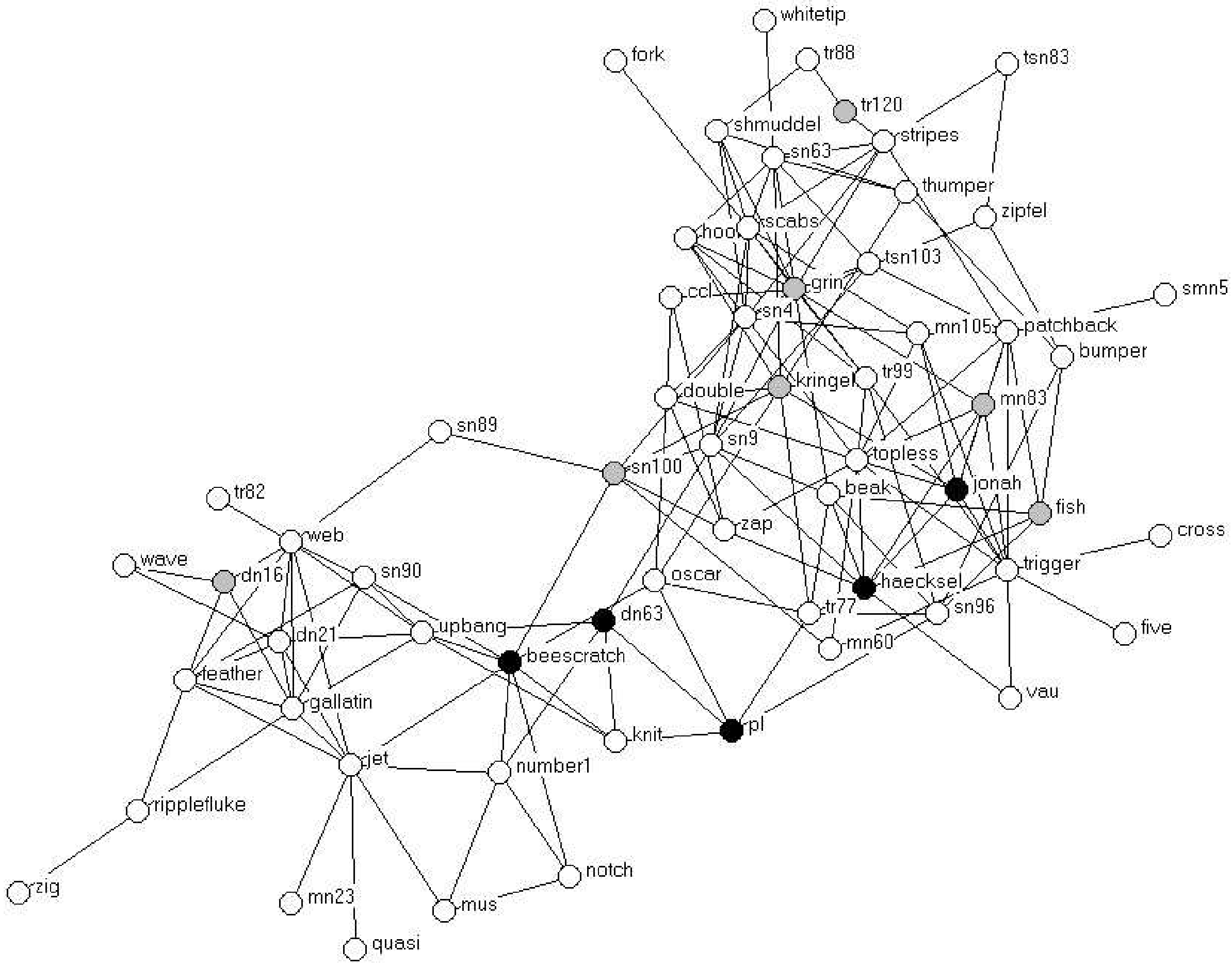

**Figure 1.** Social network of bottlenose dolphins in Doubtful Sound, New Zealand; each vertex represents an individual and each edge represents a pair that was observed in the same school more often than expected by chance; see (*26*) for more details on how the social network was constructed. Dolphins observed side flopping (SF) are in black and the ones observed upside-down lobtailing (ULT) are in grey.

# Results

During the study period I spent 137 days (879.2 hours) looking for dolphins. I followed focal groups for 716.5 hours (over 133 days). During this time I was able to identify reliably the identity of individuals performing side flops in 10 instances and performing upside-down lobtails in 15 cases. Most side flops were performed by males (9 out of 10). The likelihood that 9 out of 10 SF were performed by males and that females and males had equal chances to perform them was very low (p=0.001, using a randomisation test with 1000 iterations). In contrast, most ULT were performed by females (14 out of 15) and the likelihood that males and females were equally likely to perform them was also low (p=0.002, 1000 iterations).

All individuals were equally likely to be recognised when performing SF or ULT because of the distinct markings individuals bear on their dorsal fins. Only five males were identified performing the 10 SF (Figure 1). A randomisation test (10000 iterations) showed that the likelihood that all males in the population were equally likely to perform SF was low (p=0.0006, likelihood that 10 SF were observed and 5 out of all males were identified performing them given the amount of time we spent observing each of them). Not all females seemed to perform ULT either (Figure 1).

Only seven females were identified performing the 15 ULT which is highly unlikely if all females were equally likely to perform this behaviour (p=0.003, 1000 iterations randomisation test: likelihood that 15 ULT are performed by seven females when drawn randomly from all the females in the population given the amount of time we spent observing each of them).

These tests show that SF tends to be a male-specific behavioural event and ULT a female-specific one. It is worth noting that both the SF performed by the female and the ULT performed by the male were not followed by changes in the school's behavioural state. In addition not all individuals seem to use these signals in the population. I therefore tested whether males that performed SF and females that performed ULT tended to have higher centrality measures in the social network.

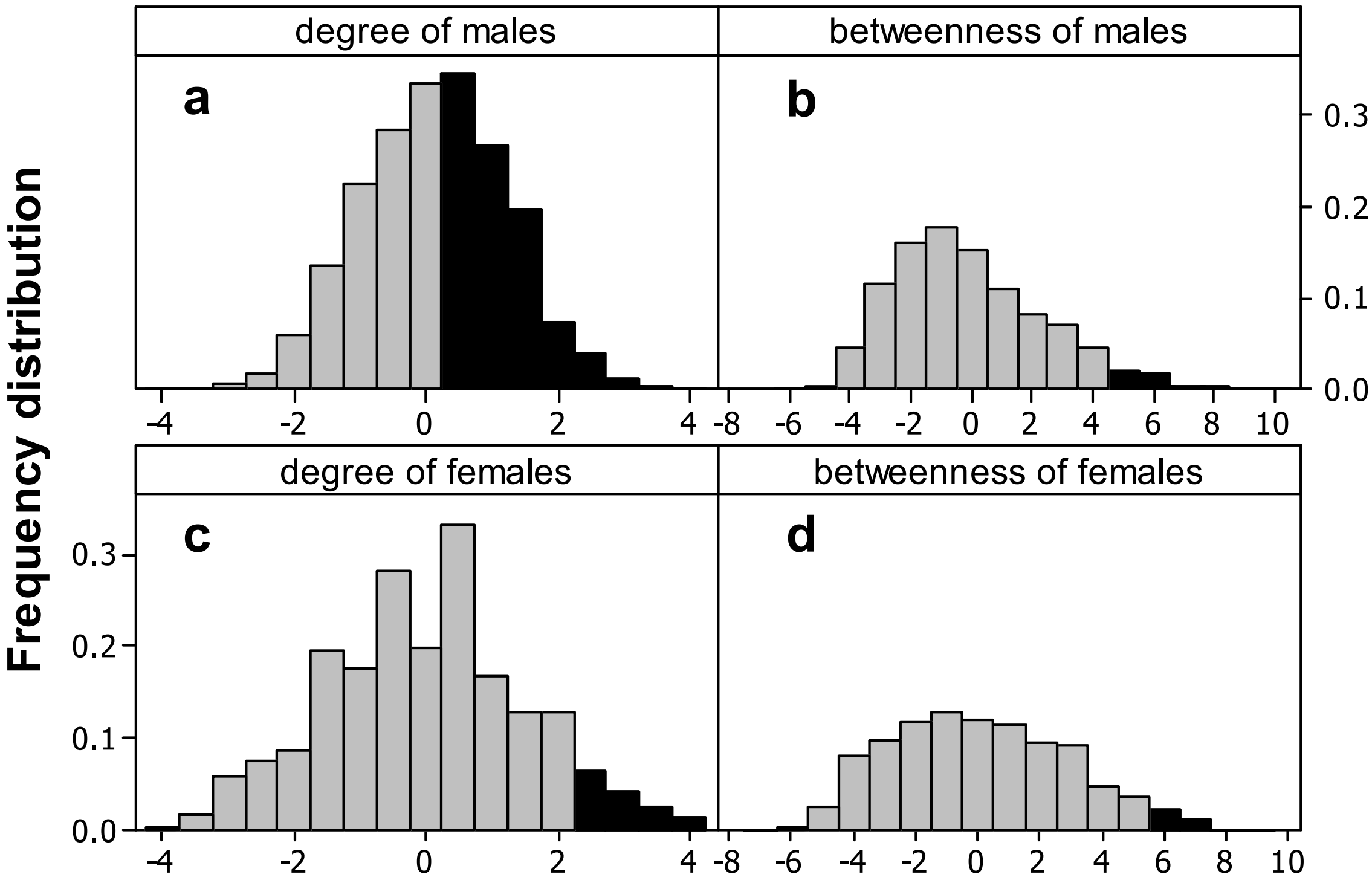

**Figure 2.** Results of the four weighted randomisation tests looking at the difference in betweenness (a and c) and degree (b and d) of individual males observed side flopping versus others (a and b) and individual females observed upside-down lobtailing versus others (c and d). 1000 randomisations were performed for each test and each panel is a histogram of the resulting difference in degree and betweenness at each randomisation. In this analysis the likelihood that an individual was observed performing a behaviour in the random data was weighted by the amount of time spent with that individual in the field. The values that were higher than the observed differences are shown in black bars.

The difference in degree, the number of partners an individual had on the social network, between SF males and non-SF males was small (0.29) and did not differ from random differences obtained by randomising who had been observed performing SF (1000 iterations, p= 0.382, Figure 2b). Similarly ULT females did not have a significantly higher degree than non-ULT females (difference=2.58, 1000 iterations, p=0.056, Figure 2d). However, both SF males and ULT females had significantly higher betweenness values, a measure of the diversity of links an individual had, than non-SF males and non-ULT females respectively (males: difference = 4.50, 1000 iterations, p = 0.041, Figure 2a; females: difference = 5.20, 1000 iterations, p = 0.040, Figure 2c). These tests were ran 100 times to test the power of the randomisations and the same level of significance (p>0.05 for degree and p<0.05 for betweenness) was obtained in all 100 runs except for the female degree test which was significant in 2 instances. The degree and betweenness of both males and females were weakly correlated (Pearson's r: r=0.46 and p=0.021 and r=0.58 and p=0.003 respectively).

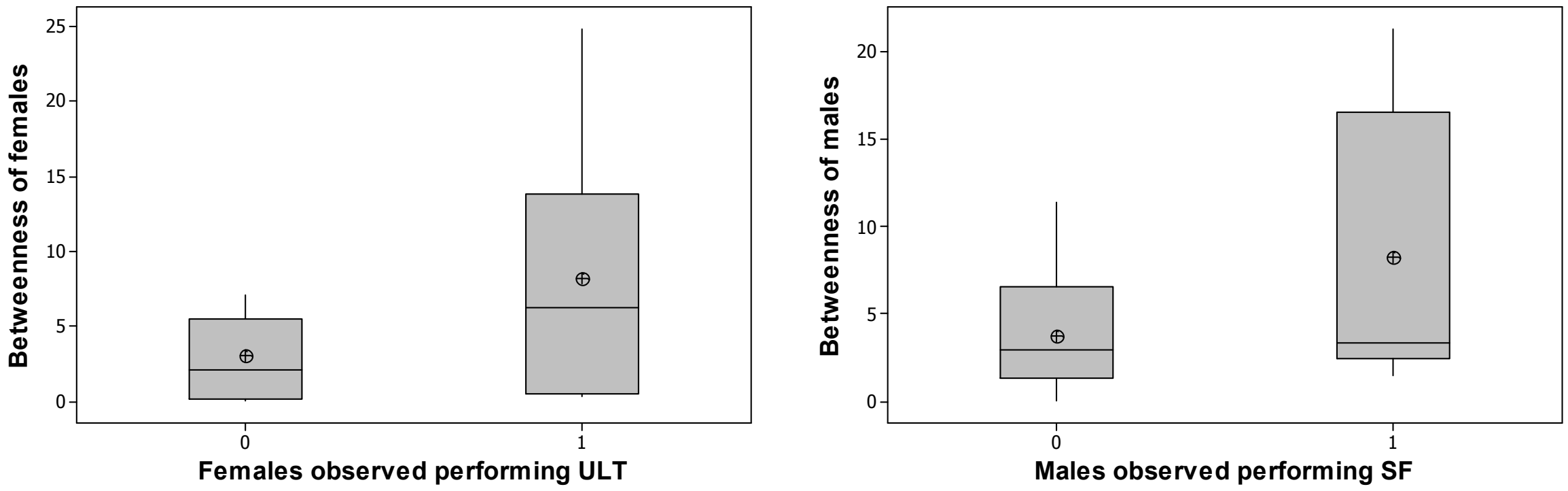

**Figure 3.** Boxplot of betweenness for females (a) and males (b) depending on whether they were observed performing ULT or SF (value 1 on x-axis) or not (value 0 on x-axis). Lines in boxes are medians and symbols are means.

# Discussion

This study provides some initial evidence on the decision making processes, and the mechanisms involved to reach consensus, in situations where information is not equally shared by all individuals in fission-fusion societies. It appears that in this population unshared decision making is used to synchronise the activity of school members using a signal that allow global communication at the school level but is advantageous in that it prevents advertisement of activity shift to non-school members.

SF and ULT are rare events and therefore could lack the necessary reinforcement to be a useful signal in the population. However, recent studies (*48*) show that rare events can still contribute to maintaining stable resource networks. It is possible that these events are only used when vocalisation is not possible because, for example, direct competitors are close by and this situation only arises rarely. Indeed, vocalisations can be heard up to several kilometres from the emitter and it might therefore be difficult to control to whom these vocalisations are advertised (*49*). The sound produced by percussive behaviours such as side flops and upside-down lobtails do not propagate more than a few 100 meters and would therefore not be heard by individuals that are further away (*31*). There is circumstantial evidence that schools are rarely in the vicinity of one another in Doubtful Sound (*24*) but further studies are needed to confirm whether SF and ULT are more likely to occur when schools are close to one another.

Bottlenose dolphins in Doubtful Sound rely on reef-associated prey items that are produced within the fjord system and are spatially patchy but fixed (*50*). These prey items tend to have a slow growth and there are good evidence that the dolphin population's carrying capacity is limited by food (*50*). Intra-specific competition for food therefore plays an important role in defining the fitness of individuals which could explain the evolution of the signalling system described here. Complex social behaviour have been described arising from selective forces in other bottlenose dolphin populations (*51*). Centrality measures are not related to age or sex in this population (*27*). We are therefore left with the hypothesis that the behavioural role highlighted by this study is not associated with the individual but with its position in the social network, potentially indicating a social role (*52*). This hypothesis will be difficult to test empirically in the field because of the ethical issues surrounding playback and knockout experiments on free-ranging cetaceans. However, natural experiments, through the natural disappearance of individuals, could be helpful in testing this theory.

Signals associated with shift towards travelling were performed by males while signals associated with the ending of travelling bouts were carried out by females. The Doubtful Sound population live in mixed-sex school year-round which is unique for this species and rare for a fission-fusion society (*24*). The reasons for this are still largely unknown but may include long-term mate guarding, infanticide avoidance, long-term paternal role in young rearing, or lack of dispersal from the maternal unit (*24*). While sexual dimorphism is not pronounced in bottlenose dolphins (*53*), some difference in metabolic costs still exist (*37, 54*) and cost of transport tends to be more expensive for females. These differences are enhanced by the extreme location of Doubtful Sound for the species (the population live at the southern limit of the species' range). This sexual discrimination in signal production could therefore be explained by the optimisation of the cost of transport for individuals allowing mixed-sex school to remain synchronous.

Individuals with high betweenness values in principle will have had more diverse affiliations within the social network and hence will be more knowledgeable about potential competitors because they have been more exposed to them. In addition, they may as well have knowledge about the patches recently visited by those and therefore would have a better understanding of the current quality of food patches. That is they would be more likely to know which food patches have been visited by those other groups and therefore would know which ones to avoid. Dietary analyses show that the Doubtful Sound population of bottlenose dolphins rely on reef-associated prey items which are resident to the fiord (*50*). Therefore, having some understanding of which of those patches have been visited by other groups in the recent past provides a good proxy for patch quality. Following such individuals would be advantageous in decreasing travelling costs for all in the school. Individuals with high degree can indiscriminately reach more individuals within the network and may therefore be more related to the archetypical symbolic representation of leaders (*55, 56*). They therefore would have a good knowledge of individuals in their immediate vicinity, in their local cluster, but those can include both within and between 'global' cluster links. Following the Conradt-Roper framework (*17*) they would therefore be less reliable sources in decision making processes regarding competition avoidance (scrambled or direct). The relative relationship between these two network statistics can explain the marginally similar results for degree and betweenness for females. However, for both males and females betweenness provided more explanation of the heterogeneity of the data than degree did.

Since signallers seem to be more likely to have preferred companionships (*26*) in several clusters of individuals, there can be several direct and indirect benefits for signalling. Further studies on the genetic relationship between the signallers and the members of the schools in which these individuals are observed signalling could help tease apart the roles of inclusive fitness and cooperation (*57-59*) in the evolution of these signals.

## Acknowledgements

I am currently supported by a Killam Postdoctoral Fellowship provided by the Killam trusts. I would like to thank Ramon Ferrer-i-Cancho and Hal Whitehead for numerous fruitful discussions. Comments from two anonymous reviewers and Sara Helms Cahan improved this manuscript. Data collection and compilation was funded by the New Zealand Whale and Dolphin Trust, the New Zealand Department of Conservation, Real Journeys Ltd, and the University of Otago (Departments of Zoology and Marine Sciences and Bridging Grant scheme). I would also like to thank Susan M. Lusseau, Oliver J. Boisseau, Liz Slooten, and Steve Dawson for their numerous contributions to this research.